# NIR/VIS dual-comb spectroscopy comparing high and low repetition rate regimes

Alexander Eber[1,†], Mithun Pal[1,†], Lukas Fürst[1], Emily Hruska[1], Christoph Gruber[1], Marcus Ossiander[1,2] and Birgitta Bernhardt[1,*]

[1]Institute of Experimental Physics, Graz University of Technology, Petersgasse 16, 8010 Graz, Austria

[2]Harvard John A. Paulson School of Engineering and Applied Sciences, 9 Oxford Street Cambridge, Massachusetts 02138, United States

*bernhardt@tugraz.at

[†]contributed equally

## Abstract

Dual-comb spectroscopy enables broadband analysis of key molecules with unparalleled frequency resolution and exceptional signal-to-noise ratios across various spectral regions. However, fully harnessing its potential for broadband spectroscopy with high sensitivity and spectral resolution depends critically on selecting the appropriate frequency combs with optimized (comb) parameters tailored to specific applications. This study compares dual-comb spectroscopy systems operating at 80 MHz and 1 GHz repetition rates, in the near infrared and visible spectral regions. The 80 MHz system provides high spectral resolution, ideal for resolving complex spectra, showcased with measurements of $NH_3$ vibrational bands and $I_2$ hyperfine transitions. Utilizing phase-locked feed-forward stabilization, the system delivers excellent signal-to-noise ratios but faces limitations in temporal resolution. The free-running 1 GHz system offers superior temporal resolution and compactness, making it suitable for real-time environmental monitoring in laboratory and field settings. A self-correction algorithm advances the high mutual coherence, enabling high-signal-to-noise measurements without additional electronics. With its 1 GHz resolution, it excels in monitoring $NH_3$ transitions or $NO_2$ lines at high speeds. This work highlights the complementary strengths of these systems for high-resolution spectroscopy and real time trace gas sensing.

# Introduction

Dual-comb spectroscopy (DCS) has emerged as a ground-breaking spectroscopic technique, offering unparalleled potential across a diverse range of applications [1,2]. The realization of DCS relies on an interferometric principle with asynchronous optical sampling, where the interference of two optical frequency combs (OFCs) with slightly different repetition rates ($f_{rep}$ and $f_{rep}+\Delta f_{rep}$) is recorded in the time domain. By performing a fast Fourier transform (FFT) of the time signal, the broadband laser spectrum is obtained. Unlike traditional Michelson-based Fourier-transform spectrometers, where the required mechanical scanning time and range can be a limiting factor, DCS achieves down-conversion of all frequencies within one interferogram. The resulting radio frequencies can be detected by a simple photodetector. This down-conversion is comparable to Fourier Transform Spectroscopy (FTS); however, it is accomplished more efficiently (i.e., with higher down-conversion factors). Additionally, the difference in repetition rates between the two combs allows for a static measurement without relying on moving components. As a result, DCS exhibits superior temporal resolution compared to conventional FTS. Furthermore, DCS allows absolute frequency calibration with high accuracy, traceable to atomic clocks, which ensures precision and reliability. The combination of these key advantages positions DCS as a transformative tool for high-resolution spectroscopy and advanced investigations in molecular spectroscopy [3–7], environmental sensing [8–10], biological and chemical sensing [11], and hyperspectral imaging [12].

However, the exploitation of the full potential of DCS critically depends on tailoring two crucial parameters with respect to the requirements of the diverse applications: the repetition rate, and with that, the possible detuning frequency, and the phase stability between the frequency combs.

The repetition rates of the combs limit the spectral resolution of the dual-comb spectrometer without adding additional complications, which enable interleaving [13]. Sustaining mutual coherence over extended timescales is essential for coherent time-domain averaging, which enables high-resolution spectra with excellent signal-to-noise ratios (SNR). Coherent averaging enhances the sensitivity and solves the secondary challenge of managing large datasets, which can overload acquisition devices [14]. The selection of the laser system, its stabilization, and the properties of the gain medium are crucial in achieving mutual coherence. Various phase measurements and locking techniques, e.g., f-2f interferometry [15], are employed to stabilize the carrier-envelope offset frequency of frequency comb sources. For DCS, it suffices to control the relative phase of the frequency combs [16]. For detuning of the repetition rates in the kHz domain, the mutual coherence between two combs can be reconstructed with post-processing algorithms [17,18].

A high pulse energy and a broad spectrum make ytterbium-doped frequency combs particularly suitable for extending the spectroscopic capabilities into the visible (VIS) and ultraviolet (UV) spectral regions, i.e., using nonlinear frequency up-conversion with nonlinear crystals or high-harmonic generation [7,19–21]. The majority of the trace gasses, e.g., $NO_2$, $O_3$, $NO_3$, $HCHO$, exhibit strong ro-vibrational transitions in the UV/VIS region, making this spectral range ideal for detecting ultra-low concentrations, often in the ppb to ppt range [22–24]. Furthermore, the absorption spectra of these gases exhibit many narrow absorption lines, making advanced spectral resolution an inevitable requirement. By employing a secondary molecular absorption signal for frequency calibration post-correction methods allow coherent averaging in the green spectral region for more than 1 s [10]. The lasers' fundamental emission in the near infrared (NIR) (~1.0 µm), where many molecules of atmospheric and spectroscopic interest (e.g., $NH_3$, $CH_4$, $CO_2$, etc.) exhibit moderate or weak overtone bands, has been exploited less than the "fingerprint" region in the MIR. However, various stabilization techniques [16], post-processing algorithms [3,25], and self-correction algorithms [17,18,26] open novel opportunities to leverage the full potential of the NIR combs situated within a less spectroscopically favorable region.

The selection of the repetition rate in traditional DCS systems introduces a trade-off between spectral resolution and acquisition speed. Lasers with MHz repetition rates can achieve spectral resolutions on this scale, making them suitable for studying molecules with MHz-order Doppler linewidths. However, the maximum detuning frequency, which determines the acquisition time, is typically limited to the sub-kHz range in such systems, thereby restricting the temporal resolution. Conversely, GHz repetition rate DCS systems excel in temporal resolution, with detuning frequencies in the tens of kHz region. This enables extensive averaging and fast data acquisition, at the expense of spectral resolution. Importantly, light atmospheric molecules exhibit GHz-order Doppler widths at room temperature and ambient pressure, making GHz-based systems well-suited for atmospheric condition studies. Recent advancements, such as single-cavity free-running GHz dual-comb systems, have significantly enhanced the robustness, compactness, and cost-effectiveness of DCS setups [17,27–30]. The single-cavity architecture inherently improves mutual coherence times by leveraging common-mode noise suppression, as both combs share the same physical environment and pump diode [31]. However, this improvement is limited over extended time scales due to relative timing jitter between the two carrier-envelope offset frequencies, leading to comb line broadening and impairing further averaging. To address this challenge, digital phase correction techniques are employed [18,32]. These methods utilize post-processing or self-correction algorithms to reconstruct mutual coherence between the two combs, providing an easily applicable solution without relying on complex locking schemes. These innovations pave the way for the next generation of DCS, enabling precise, high-resolution spectroscopy across all wavelengths. Additionally, the enhanced performance and portability could make DCS with GHz repetition rates an indispensable field-deployable tool for trace gas detection and molecular spectroscopy.

This work presents the first direct comparison of dual-comb spectroscopy (DCS) systems operating at 80 MHz and 1 GHz repetition rates across both NIR and VIS spectral ranges. The performance of the systems is analyzed by applying DCS to ammonia ($NH_3$) in the NIR and to iodine ($I_2$) and nitrogen dioxide ($NO_2$) in the visible range.

The 80 MHz Yb-doped DCS system utilizes a feed-forward stabilization technique [33], incorporating an acousto-optic frequency shifter (AOFS) to enable coherent averaging. To the best of our knowledge, this work presents for the first time a high-resolution dual-comb spectrum of ammonia ($NH_3$) near 1 µm, demonstrating the system's capability for high-resolution spectroscopy in the NIR. An optional second harmonic generation after stabilization allows switching between the NIR and VIS. In the visible, the 80 MHz system achieves comb-resolved dense ro-vibronic spectra of $I_2$, surpassing state-of-the-art measurements in resolution and stability over extended timescales.

The 1 GHz system utilizes a single-cavity free-running dual-comb source with four outputs, covering parts of the NIR and VIS simultaneously. This capability of concurrent dual-wavelength operation demonstrates its potential for real-time spectroscopic applications. An advanced self-correction algorithm enhances mutual coherence, enabling coherent averaging within a timescale of several seconds, currently limited by computer memory. This facilitates high-temporal-resolution detection of atmospheric trace gases with excellent SNR, ideal for environmental monitoring. Notably, this is the first demonstration of nitrogen dioxide ($NO_2$) detection using a GHz dual-comb system that combines high sensitivity with sub-second temporal resolution, highlighting its potential for real-time atmospheric monitoring.

A comparative evaluation of $NH_3$, $I_2$, and $NO_2$ spectra from both systems highlights their complementary performance across the NIR and VIS regions. Together, these results demonstrate how MHz and GHz DCS systems address different yet complementary needs, from ultra-high-resolution spectroscopy to rapid, high-sensitivity environmental monitoring.

# Materials and Methods

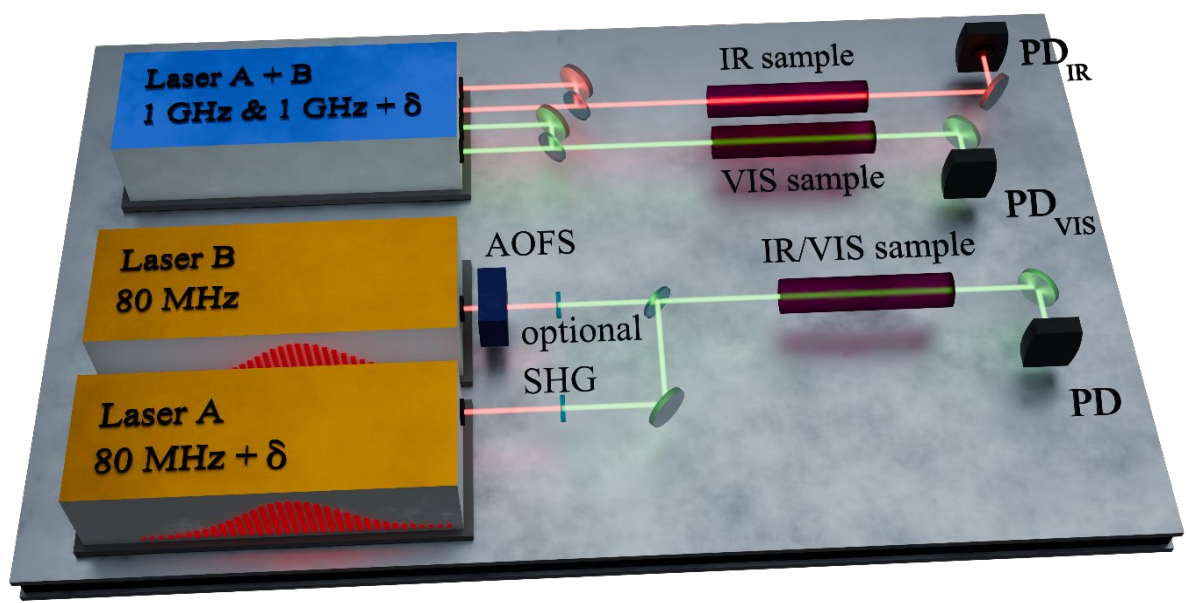

Fig. 1. **The experimental setup for the two different systems: 80 MHz repetition rate with two individual laser sources (orange boxes, bottom) and the 1 GHz repetition rate system with a single-cavity design (blue, top).** In case of the 80 MHz repetition rate system, an optional SHG scheme enables the use either the fundamental or the frequency doubled spectrum. The 1 GHz repetition rate system is capable of producing the fundamental and the frequency doubled spectrum simultaneously.

## Laser sources

Figure 1 shows the experimental setup of both 80 MHz and 1 GHz repetition rate DCS systems. While the high repetition rate system can produce the second harmonic internally, the low repetition rate system employs an optional second harmonic generation (SHG) scheme externally.

The dual-comb interferometer with 80 MHz repetition rate uses two independent ytterbium-based fiber frequency combs (orange Femtosecond Ytterbium Laser from MenloSystems) operating at repetition rates of $f_{rep,1}$ = 80 MHz and $f_{rep,2}$ = 80 MHz + 24 Hz with a pulse duration of 150 fs. The repetition rates are stabilized via piezoelectric transducers that control the cavity lengths of each comb. Both lasers are synchronized to a common 10 MHz reference, ensuring a stable repetition rate difference between the two combs. A feed-forward technique stabilizes the relative carrier-envelope offset frequency to enable coherent averaging. Details can be found in our recent work [33]. The stabilization is performed in the NIR spectral region with subsequent SHG in LBO crystals. The relative resolving power of the 80 MHz system in the VIS is $7.3 \times 10^6$ and half that in the NIR. For the experiments, an average power of 5 mW (NIR) and 1 mW (VIS) are used. The mutual coherence between both combs is maintained after the non-linear optical process.

For the 1 GHz repetition rate DCS system, a single-cavity dual-comb mode-locked Yb:CALGO laser (K2 Photonics) is employed. This system utilizes a multimode diode-pumped solid-state laser configuration with a semiconductor saturable absorber mirror for mode-locking [34]. The laser generates more than 2 W of average NIR power per comb and maintains pulse durations of approximately 100 fs at a repetition rate of 1.03 GHz and a central wavelength of 1055 nm. The repetition rate difference ($\Delta f_{rep}$) can be tuned from 0 to 200 kHz by adjusting the vertical position of a biprism within the cavity [34]. Visible light is generated by an included SHG stage inside the laser, producing average powers of up to 350 mW per comb with pulse durations of 80 fs and a spectrum centered around 527 nm. Coherent averaging is achieved by a self-referencing algorithm, which eliminates the need for active stabilization [35], and results in a relative resolving power of $5.8 \times 10^5$ in the VIS and half that in the NIR.

## Data acquisition

For the conventional dual-cavity DCS system, the VIS and NIR interferograms are recorded by a moderately fast photodetector (Thorlabs PDA10A2, 150 MHz) combined with a (5 - 48) MHz bandpass filter to avoid aliasing effects and reduce low-frequency noise. A 16-bit digitizer card with a sampling rate of 250 MSa/s is used for acquiring and storing data. Real-time averaging of interferograms (IGMs) increases the SNR and therefore the sensitivity of the system.

The 1 GHz system requires a high-speed photodetector (QUBIG GmbH) with a 1 GHz bandwidth (AC). A transimpedance bandwidth of 15 kHz rejects any low-frequency noise. An additional 500 MHz low-pass filter mitigates aliasing effects and discards the repetition rate. The time dependent signal is recorded and streamed to a PC by a 16-bit digitizer card with a sampling rate of 1 GSa/s. Afterwards, a self-correction algorithm, similar to [35], reconstructs detuning and phase changes between successive IGMs. Therefore, the algorithm first corrects the time differences between consecutive IGMs across the entire recorded trace. It then compensates phase jitter between consecutive IGMs to reconstruct comb-resolved spectra. Coherent averaging enhances the spectral resolution and improves the signal-to-noise ratio for precise spectroscopic measurements.

# Results

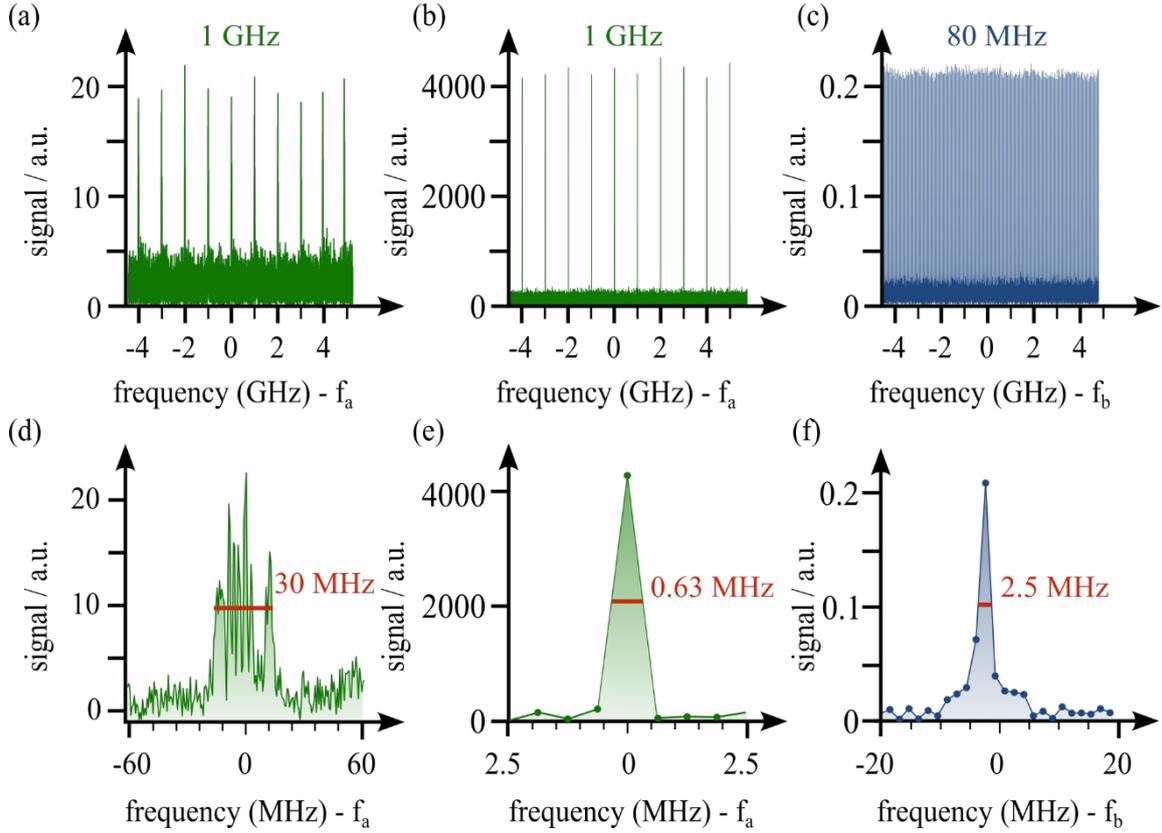

Fig. 2. **Comb-resolved spectra of both systems in the optical frequency domain.** The frequency axis is shifted by $f_a = 285$ THz and $f_b = 290$ THz towards zero. (a) Ten uncorrected comb modes of the GHz system recorded over 20 ms with a detuning of 80 kHz. (b) Ten corrected comb modes of the GHz system with a recording time of 20 ms and a detuning of 80 kHz. (c) The same 10 GHz window as in (a) and (b) is shown for the stabilized 80 MHz system. (d) A zoom into panel (a) reveals a 30 MHz broad comb mode. (e) After correction, the comb mode consists of only a single point with a width of 0.63 MHz. (f) The 2.5 MHz broad comb mode of the stabilized 80 MHz system.

To characterize the mutual coherence in both systems for extended periods, multiple consecutive interferogram bursts are recorded. A subsequent FFT extracts tooth-resolved spectra for both the 80 MHz and 1 GHz systems.

Panels 2(a) and (d) show the comb-resolved spectra for the 1 GHz DCS system obtained from a trace of 1600 consecutive IGMs. The system operates with a detuning frequency of 80 kHz. After the phase correction and resampling algorithm (Figure 2(b) and (e)), each comb line reaches an SNR of 200. We compare the inherent mutual coherence of the 1 GHz *single*-cavity DCS system with that of a *dual*-cavity 1 GHz DCS system characterized previously [36] by analyzing comb-resolved spectra without applying any corrections in both cases. In [36], the uncorrected comb line width was found to be 300 MHz for a detuning of 80 kHz and 20 ms recording time. The comb modes of the uncorrected single cavity system presented here exhibit an order of magnitude lower width at 30 MHz (Figure 2(d)) under the same experimental conditions. This finding indicates enhanced intrinsic mutual coherence with the single cavity geometry. After applying the phase correction algorithm, the comb mode width reduces to 50

Hz in the radio-frequency (RF) domain and 0.63 MHz in the optical domain, for both GHz systems, the dual-cavity system from literature [36] and our single-cavity system. This matches the Fourier-transform limit of 0.63 MHz.

Figure 2(c) presents the comb-resolved spectrum obtained using the 80 MHz DCS system with feed-forward stabilization. The detuning is 100 Hz with a recording time of 400 ms, corresponding to 40 IGMs. The measured full width at half maximum of 2.5 MHz approaches the Fourier-transform limit for a 400 ms recording time of 2 MHz. Detailed characterization of the comb spectra and the comb mode width can be found in our recent work [33]. The main limitation for averaging is the slow dynamic drift of the carrier-envelope offset frequency ($f_{ceo}$) of the individual combs, limiting the measurement time to 30 minutes. This can be extended to several hours by implementing a slow feedback loop for compensating the drifts. Due to the disparate parameter spaces of the two systems, a fair comparison with the same number of IGMs or acquisition time is not feasible due to memory constraints of the 80 MHz system.

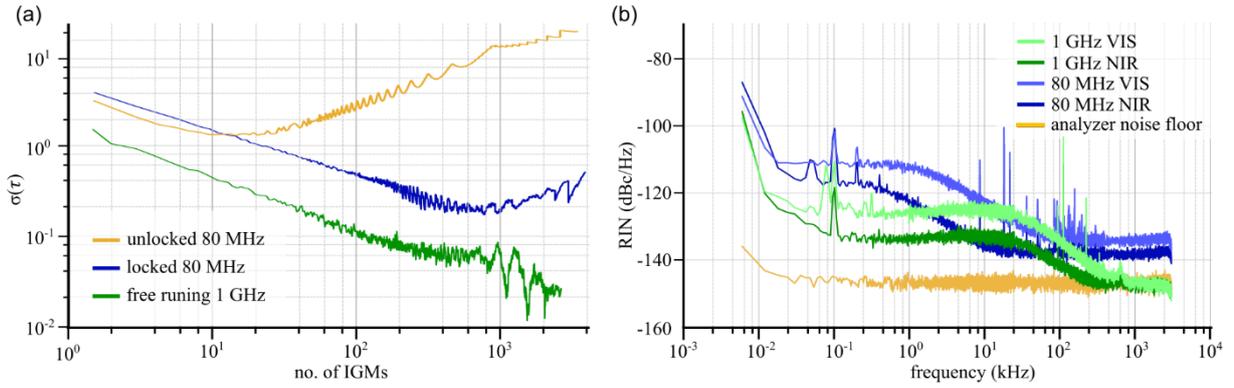

Fig. 3. (a) Allan deviation of peak frequency of the RF spectrum for the 80 MHz system in unlocked (green) and feed-forward stabilized (red) operation, compared with the free-running 1 GHz dual-comb source (blue). The data for 80 MHz system were reused from Ref. [33] with permission. Copyright 2022 American Association for the Advancement of Science. (b) Relative intensity noise (RIN) spectra of the two systems in the NIR and VIS regions, together with the analyzer noise floor. The 1 GHz system exhibits lower RIN in both spectral ranges, demonstrating its advantage for high-SNR and short-timescale detection.

We investigated the peak frequency drift of DCS spectra across consecutive IGM events for the unlocked 80 MHz system, the feed-forward stabilized 80 MHz system, and the free-running 1 GHz system using Allan variance analysis (see Figure 3a). A detailed comparison of the locked and unlocked 80 MHz operation is provided in Ref. [33]. The unlocked 80 MHz system exhibits pronounced phase fluctuations between the two combs over time, indicating that without stabilization it cannot sustain long-term coherence, thereby limiting accurate spectroscopic measurements. In contrast, the free-running 1 GHz dual-comb source (green curve) demonstrates comparatively stable phase correlation with respect to the stabilized 80 MHz system, attributed to its higher detuning frequency, which inherently reduces relative phase fluctuations. However, since its repetition rate is not locked and the comb detuning is not fully stabilized, self-phase correction remains necessary to achieve high spectral quality.

In addition, we characterized the relative intensity noise (RIN) performance of both DCS systems, as RIN is a critical parameter influencing the sensitivity of DCS. The RIN measurement results for both the 80 MHz and 1 GHz systems in the NIR and VIS spectral regions are presented in Figure 3(b). The measurements were performed using a commercial RIN analyzer (PNA1, Thorlabs). In DCS, optical noise, including RIN, at or near the detuning frequency ($\Delta f$) is down-converted into the RF domain, where it directly affects the SNR of the DCS spectra. Elevated RIN around $\Delta f$ leads to an increase in detection noise, thereby degrading single-shot sensitivity and requiring extended averaging to achieve a desirable SNR. This is particularly relevant for the 80 MHz system, where typical detuning frequencies are limited to a few hundred hertz, within the regime where 1/f noise dominates. In contrast, the 1 GHz system permits larger detuning frequencies (on the order of several kHz), enabling spectral mapping in RF regions with significantly reduced RIN. As a result, the 1 GHz system offers improved performance in terms of reduced acquisition time and lower noise-equivalent absorption (NEA).

## Near-infrared dual-comb spectroscopy

Beyond its spectroscopic importance, ammonia ($NH_3$) is a significant atmospheric trace gas, primarily as a by-product of livestock and fertilizer production [37]. Monitoring $NH_3$ in ambient air is challenging due to its weak absorption cross-section (~$10^{-22}$ $cm^2$/molecule) in the NIR spectral range and low environmental concentrations, often in the ppb range. Despite having a relatively simple polyatomic structure, $NH_3$ exhibits complex spectral characteristics [38]. This complexity arises from transitions involving combinations of two or more vibrational modes, leading to doublet and triplet features due to degenerate vibrational states [39,40]. High-resolution spectra are crucial for assigning these transitions, which, in turn, can improve the accuracy of electronic-structure calculations via ab-initio methods.

The performance of both coherent DCS systems in the NIR spectral range is characterized by monitoring the absorption spectra of ammonia ($NH_3$). For the 80 MHz system, 2300 individual 42-ms-long IGM traces are averaged over a total duration of 180 seconds. The experiment utilizes an interaction path length of 105 cm, with 800 mbar $NH_3$ at 298 K. After performing a FFT and up-converting the frequency, a broadband spectrum covering 9 THz is obtained. Figure 3 (blue line) shows the $NH_3$ absorption spectrum with a resolution of 80 MHz. The spectrum exhibits a SNR of above 1000, which can be attributed to the stabilization and coherent averaging method, resulting in a noise equivalent absorption (NEA) of $1 \times 10^{-5}$ $cm^{-1}$ at 180 s, and aligns well with HITRAN data. To our knowledge, the resolution achieved in this work surpasses previous studies using FTIR or dual-comb spectrometers near wavelengths of 1.0 μm [38,40]. The 80 MHz system, while offering high resolution, faces limitations due to its low detuning frequency, resulting in insufficient SNR for short timescales. Additionally, the feed-forward stabilization approach involving a continuous-wave (CW) laser, two beat detection schemes, and phase-locked loops is less practical for field-deployable spectrometers, necessitating alternative simplified configurations.

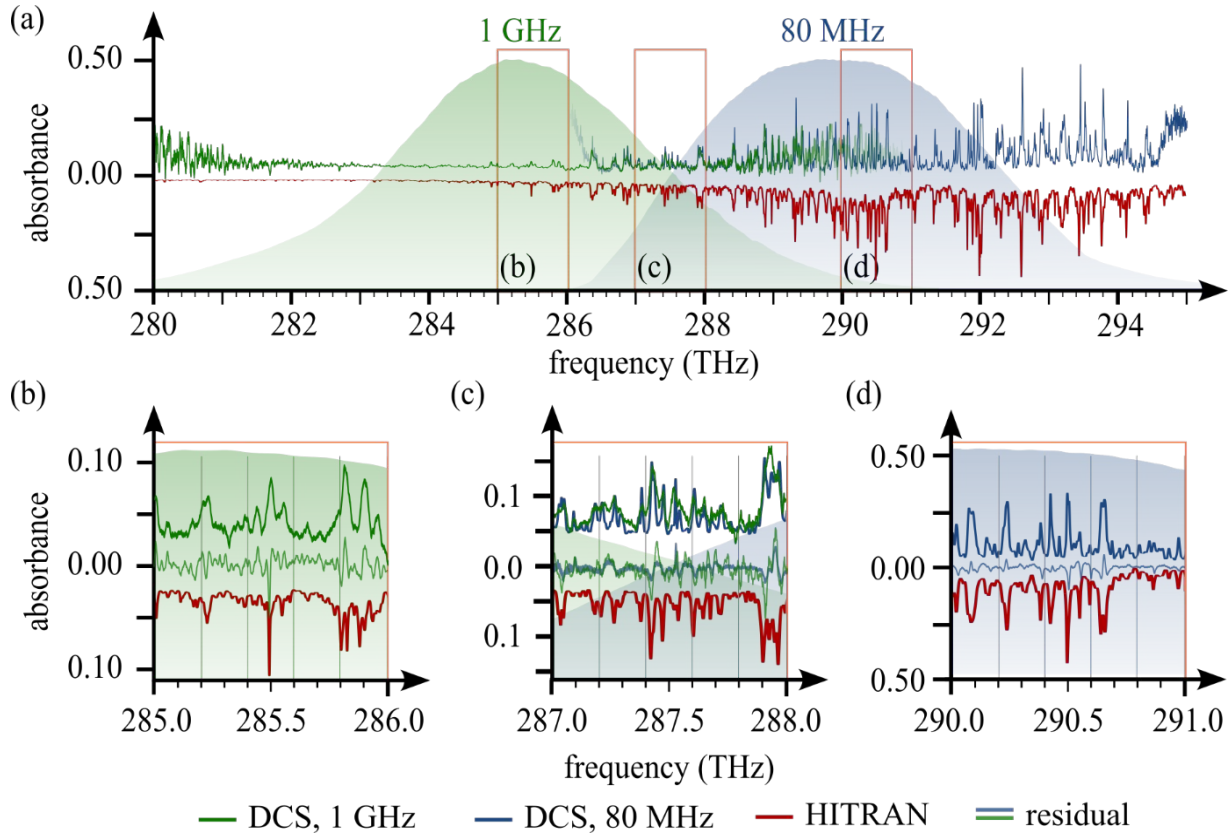

Fig. 4. **The absorbance of 2 300 ammonia traces from the 1 GHz and the 80 MHz systems compared to HITRAN** [41,42]. (a) The spectrum of each system (1 GHz system: green, 80 MHz system: blue) covers more than 8 THz. The individual laser spectra are shaded in the background and a simulated HITRAN [41,42] reference (red line) is provided. (b) A zoom into the 1 GHz system. Panel (c) shows a part of the spectral region in which both lasers have significant spectral intensity and the absorbance traces overlap, providing a comparison between both systems. (d) The high-resolution 80 MHz spectrum is compared to literature, resolving the complex absorption structure of the ammonia in this spectral region. The residual between the measurements and HITRAN is shown in light green/blue.

We evaluate the 1 GHz system by measuring the same $NH_3$ absorption spectra under identical experimental conditions. With a recording time of 115 ms, 2300 IGMs are averaged for this experiment followed by a posteriori phase correction. Figure 4 (green line) shows the $NH_3$ absorption features measured with the 1 GHz system. The achieved spectral resolution of 1 GHz is adequate for monitoring $NH_3$ in ambient air while insufficient for high-resolution spectroscopy. At atmospheric pressures, pressure broadening dominates (typical $NH_3$ widths of 3 GHz [43,44]), making this resolution sufficient to capture relevant spectral features. Additionally, the high detuning frequency, reaching several tens of kHz, enables the averaging of a larger number of IGMs within a fraction of the measurement time required for the 80 MHz measurements presented above. This approach combines high SNR values with a high temporal resolution. The current setup achieves a NEA of $5 \times 10^{-5}$ cm$^{-1}$ at 500 ms. This performance can be enhanced by three orders of magnitude through the implementation of high-responsivity detectors, extended averaging times, and the integration of cavity-enhancement techniques [38]. However, the performance demonstrated here proves that a compact GHz DCS with self-correction can detect weak overtone bands without complex stabilization procedures or sophisticated cavity enhancement.

## Visible dual-comb spectroscopy

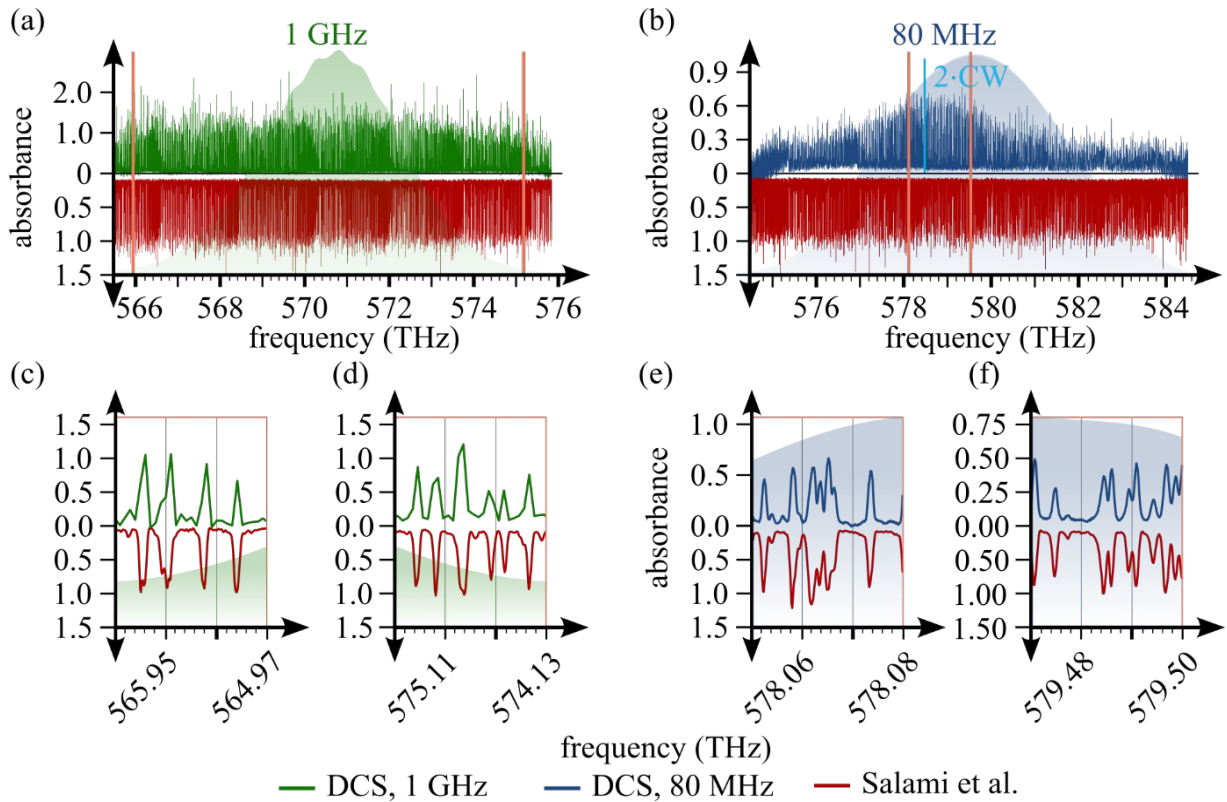

Fig. 5. **Iodine spectroscopy comparing two different DCS systems with a 20 cm long iodine cell at room temperature.** The experimental data (green/blue curves) are compared to Salami et al. [45] (red curves). (a) The complete absorbance spectrum was measured with the free-running 1 GHz laser system. (b) The entire absorbance spectrum was obtained by the relative-phase-stabilized 80 MHz fiber lasers. The position of the CW laser, which is used to stabilize the carrier envelope offset frequency, is marked in bright blue. (c) and (d) show zoom-ins into the edges of the 1 GHz absorbance spectrum of (a). (e) and (f) show zoom-ins, which reveal the individual absorption lines of iodine, measured with 80 MHz resolution compared to literature [45].

We investigate the performance of the two DCS systems in the visible spectral range. Figure 5 shows the iodine absorbance over 10 THz measured with both systems. Figure 4(a) depicts the iodine spectrum obtained with the 1 GHz system, compared to literature data. The results are obtained by coherently averaging 9900 individual IGMs over 0.5 s. More than 1700 iodine lines are resolved across the whole spectrum. While the moderate resolution of 1 GHz does not resolve all hyperfine transitions of iodine, Figures 4(c) and (d) show good agreement of the main absorption resonances. Figure 5(b) illustrates the iodine absorption spectrum obtained from the stabilized 80 MHz DCS system at around 580 THz. A detuning frequency of 24 Hz results in 42 ms acquisition time for one interferogram. The iodine in the sample cell with a length of 20 cm has a vapor pressure of 0.2 mbar at room temperature. Coherently averaging over 2600 interferograms significantly enhances the SNR and enables full utilization of the spectrometer's potential with a resolution of 80 MHz (see Figure 5(e) and (f)).

The measured spectra, centered on the B-X ro-vibronic transitions of $I_2$ [46], are compared to the ASCII iodine atlas obtained from Michelson-based Fourier transform spectroscopy (FTS) with a spectral resolution of 600 MHz [45]. The coherently averaged iodine spectrum is in excellent agreement with the iodine atlas across a spectral span of 4 THz around the continuous-wave laser. The densely packed iodine spectrum, containing hyperfine-split lines, is resolved with the accessible resolution of 80 MHz, resulting in more than 1500 detected iodine absorption lines across the spectrum. Due to the dense spectral grid in the visible range, iodine has long been used as a frequency standard for high-resolution spectroscopy and calibration purposes. The phase-locked feed-forward scheme proves its performance even when nonlinear frequency up-conversion is involved. In comparison to the work by Salami et al. [45], which achieved an SNR of 55 around 579 THz over an 8-hour timescale, our approach delivers twice the SNR within a 300-fold shorter timeframe. Additionally, we achieve a spectral resolution of 80 MHz in just 42 ms, which represents a $6.6 \times 10^6$ times improvement in acquisition speed. This highlights the spectral resolution and temporal efficiency enabled by combining DCS with our feed-forward stabilization technique.

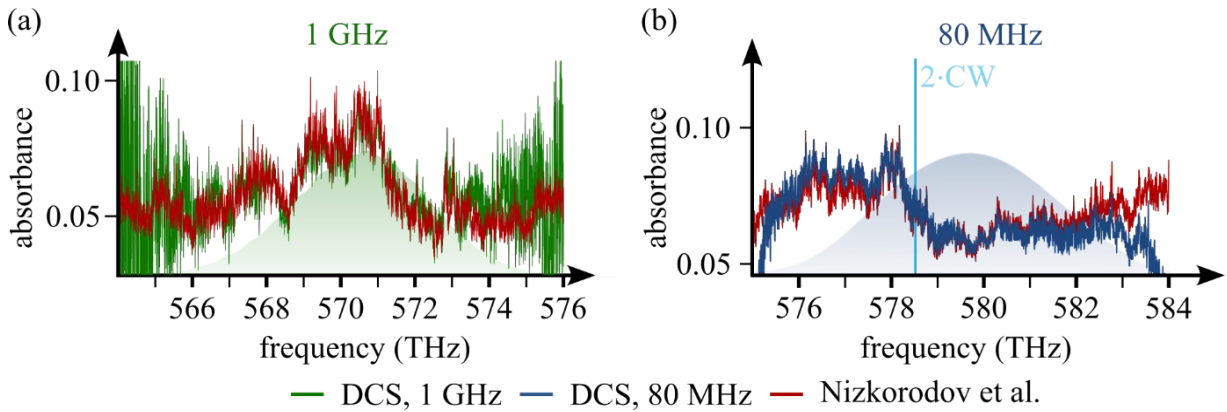

Fig. 6. **Nitrogen dioxide absorption spectroscopy with two different systems.** The experimental data (green and blue curves, respectively) are compared to Nizkorodov et al. [47] (red curves). (a) Comparison of the 1 GHz system measured with a detuning of 19.8 kHz and a total acquisition time of 500 ms. (b) Measurement with the phase-locked 80 MHz system with 24 Hz detuning and averaged over 2500 interference patterns, resulting in a total acquisition time of 100 s.

Additionally, the GHz system's capability for environmental trace gas sensing is investigated by monitoring $NO_2$ transitions. Figure 6(a) shows the $NO_2$ spectrum of the 1 GHz system in the VIS, recorded with a detuning frequency of 19.8 kHz and a measurement time of 0.5 s. The Doppler width of most $NO_2$ transitions is approximately 1 GHz at room temperature [47], making the resolution adequate for atmospheric sensing. The experiment is conducted at a gas cell pressure of 700 mbar, utilizing a multipass configuration that achieves an effective optical path length of 245 cm by folding the path seven times in a 35 cm-long cell. Currently, an NEA of $1.25 \times 10^{-5}$ cm$^{-1}$ is achieved within a 0.5 s integration time, demonstrating the potential for fast, high sensitivity measurements. This performance indicates that for field measurements with optical path lengths typically spanning several kilometers, the sensitivity could improve by three orders of magnitude. Such advancements would enable precise monitoring of atmospheric $NO_2$ at real-world concentrations on the low ppb level with high temporal resolution, offering significant benefits for real-time environmental sensing and air quality assessment. Figure 6(b) demonstrates excellent agreement between the 80 MHz system, with a

24 Hz detuning frequency, and literature. However, observing GHz-scale Doppler-broadened transitions with an 80 MHz resolution is not necessary for trace gas sensing in ambient conditions. Using a phase-locked feed-forward stabilization method, we achieve an NEA of $10^{-3}$ cm$^{-1}$ over 100 seconds by averaging 2400 interferograms. This system is less adequate for real-time trace gas sensing, where high temporal resolution and high sensitivity are essential. Furthermore, the feed-forward locking scheme is not well-suited for field-deployable setups, as it relies on acousto-optic components and feedback electronics that are susceptible to environmental perturbations, potentially affecting system robustness and reliability in non-laboratory conditions.

## Discussion and Conclusion

Table 1 summarizes our results and compares dual-comb spectroscopy systems with repetition rates of 80 MHz and 1 GHz.

Tab. 1: Overview of the experimental parameters of the 80 MHz and 1 GHz system.

|  | 1 GHz System | 80 MHz System |
|---|---:|---:|
| *Central wavelength / nm* | 1050 | 1035 |
| *Central frequency / THz* | 285.5 | 289.6 |
| *Spectral coverage / nm* | 40 | 35 |
| *Spectral coverage in the frequency domain / THz* | 11 | 10 |
| *Repetition rate detuning / Hz* | $1 - 80 \times 10^3$ | 1-200 |
| *Measurement time / s* | 0.5 | 100 |
| *Number of averages / no. of IGMs* | 10000 | 2500 |
| *NEA IR / cm$^{-1}$ (corresponding measurement time)* | $5 \times 10^{-5}$ (0.5 s) | $1 \times 10^{-5}$ (180 s) |
| *NEA VIS / cm$^{-1}$ (corresponding measurement time)* | $1.2 \times 10^{-5}$ (0.5 s) | $\sim 10^{-3}$ (100 s) |
| *DCS quality factor / $\sqrt{Hz}$* | $3.1 \times 10^6$ | $12.5 \times 10^6$ |

The 80 MHz system achieves an exceptional spectral resolution (80 MHz) with a phase-locked feed-forward stabilization technique, pinpointing fine ro-vibrational structures in $NH_3$ and $I_2$ that are crucial for detailed molecular studies and precise frequency calibration. It excels in high-resolution molecular spectroscopy by providing unparalleled resolution. However, a dependence on intricate instrumentation, minutes of measurement time, and mechanical instabilities limit the suitability for ad hoc applications such as field sensing or real time tracking of gas concentration changes.

In contrast, the 1 GHz DCS system with sub-second measurement times is well suited for novel environmental field applications such as monitoring $NO_2$ and $NH_3$ dynamics in complex urban or industrial scenarios. The single cavity design combined with self-referencing algorithms enhances portability and facilitates short measurement times on the sub-second scale with high sensitivity, critical for real-time environmental monitoring, while retaining a resolution more than sufficient to investigate broadened lines in the atmosphere.

Together, these complementary spectrometers provide a versatile platform for diverse applications, from precise molecular spectroscopy to real-time environmental monitoring. Both systems achieve high quality factors $QF = \text{SNR} \times \frac{\Delta f}{\delta f \cdot \sqrt{T}}$ with the spectral bandwidth $\Delta f$, the spectral resolution $\delta f$, and the acquisition time $T$ [48]. The 1 GHz system achieves $3.1 \times 10^6 \sqrt{Hz}$, while the 80 MHz platform reaches $12.5 \times 10^6 \sqrt{Hz}$. The comparative DCS work presented here not only advances the state-of-the-art in spectroscopy but also opens new pathways for addressing scientific and environmental challenges, such as high-precision molecular spectroscopy and real-time atmospheric monitoring with unprecedented accuracy.

## Acknowledgements

The authors gratefully acknowledge useful discussions with Jerome Genest. The authors also thank Arnaud Mussot for providing the RIN measurement device.

## Funding

The authors gratefully acknowledge support from NAWI Graz. B.B. acknowledges funding from the European Union (ERC HORIZON EUROPE 947288 ELFIS) and the Austrian Science Fund (FWF) [10.55776/Y1254]. For open access purposes, the author has applied a CC BY public copyright license to any author-accepted manuscript version arising from this submission. M.O. acknowledges funding from the European Union (grant agreement 101076933 EUVORAM). The views and opinions expressed are, however, those of the author(s) only and do not necessarily reflect those of the European Union or the European Research Council Executive Agency. Neither the European Union nor the granting authority can be held responsible for them.

## Data availability

Data underlying the results presented in this paper are available from the authors upon reasonable request.